\documentclass[showpacs,showkeys,12pt]{revtex4}%
\usepackage{amssymb}
\usepackage{amsmath}
\usepackage{amsfonts}
\usepackage{graphicx}
\usepackage{ulem} %\sout{strike out}
\usepackage[usenames]{color}

\providecommand{\av}[1]{\left\langle #1 \right\rangle} %
\providecommand{\rbr}[1]{\left( #1 \right)}%
\providecommand{\sqbr}[1]{\left[ #1 \right]} %
\providecommand{\mtiny}[1]{\text{\tiny{#1}}}%
\def\ra{\rightarrow}

\def\kB{k_{\text{\tiny{B}}}}

\begin{document}

\title[ ]{Clausius vs. Boltzmann-Gibbs entropies}
\author{Thomas Oikonomou}
\email{thoik@physics.uoc.gr}
\affiliation{Department of Physics, University of Crete, 71003
Heraklion, (Hellas) Greece}

\affiliation{Department of Physics, Faculty of Science, Ege University, 35100 Izmir, Turkey}


\begin{abstract}
In this work, we demonstrate the inappropriateness of the Boltzmann-Gibbs log-formulation of the physical Clausius entropy $S$ in connecting thermodynamics and phase space statistics. To achieve our goal, we study thermodynamically the simple case of ideal gases embedded in a finite heat bath and compare the derived equations with the phase space statistical ones obtained within the canonical ensemble. We then show that the results of the aforementioned approaches contradict to each other even in the dilute gas limit (infinite heat bath) if we request the identification of the Boltmann-Gibbs formula $\ln(Density\,\, of\,\, States)$ with the thermodynamic entropy $S$.
\end{abstract}
\eid{ }
\date{\today }
\startpage{1}
\endpage{1}
\maketitle

%\newpage
%==================================================
\section{Introduction}
%==================================================
%
The great success of the kinetic gas theory, which has essentially contributed to the  acceptance of the particle nature of the microcosmos, was the connection between the macroscopic thermodynamic observables pressure $P$ and volume $V$ (of the container/ heat bath) with the statistical mean kinetic energy $E\equiv \overline{ E_\mtiny{k}}=N\frac{1}{2}m\overline{v^2}$ of the molecules (or particles) comprising a physical system ($N$  is the total number of molecules, each of them with mass $m$, and $\overline{v^2}$ denotes the molecules mean square velocity), under the following basic postulates \cite{Demtroeder}:   i) the molecules are considered to be very small spheres with negligible eigenvolume, ii) they are in constant random motion with isotropic distributed velocities, iii) the collisions among themselves and with the walls of the container are perfectly elastic and iv) the interactions between them are negligible.
Gases obeying the aforementioned conditions are denoted as noninteracting gases (nIaG).
Then, the analytical expression of the microscopic-macroscopic connection for nIaG is computed to be
%
%---------------------------------------------------------
\begin{equation}\label{eq:intro-01}
PV=\alpha\,E\,.
\end{equation}
%---------------------------------------------------------
%
$E$ may generally present  a temperature $T$ and volume $V$ dependence, $E=E(T,V)$, and $\alpha$ can be any positive constant \cite{Einbinder}.
Considering the specific case of nIaG with constant thermal capacity $C_V$, $C_V(T)\equiv (\partial E/\partial T)_V$, we obtain a linear dependence between energy and temperature
%
%---------------------------------------------------------
\begin{equation}\label{HeatCap}
E(T,V)=C_V T+b(V)\,.
\end{equation}
%---------------------------------------------------------
%
The combination of Eqs. \eqref{eq:intro-01} and \eqref{HeatCap} with $b(V)=0$ and the constant $C_V=nR/\alpha$, yields the equation of state of the so called dilute ideal gases \cite{Laurendeau}
%
%---------------------------------------------------------
\begin{equation}\label{eq:intro-02}
PV=nR T\,,
\end{equation}
%---------------------------------------------------------
%
where  $n$ is the number of moles and $R$ is the gas constant. 
Eq. \eqref{eq:intro-02} is experimentally observed for all types of gases in the low pressure regime \cite{Weinhold} (and for temperature values quite over the condensation temperature \cite{Windisch}).
Essentially, since $T$ is finite, the low pressure condition for a given amount of moles is equivalent to the dilute gas (DiG) limit $n/V\rightarrow0$ (or $N/V\rightarrow0$).

Eqs. \eqref{eq:intro-01}-\eqref{eq:intro-02} can be reproduced in a unified manner within the phase space statistical description of the canonical ensemble (gas in contact with a reservoir) under the identification $nR=k_\mtiny{B}N$ ($k_\mtiny{B}$ is the Boltzmann constant) for $N\ra\infty$ and the specific value $\alpha=2/3$, considering a hamiltonian comprising solely of kinetic energy terms, $H(\mathbf{p})=\sum_{i=1}^{N}(p_{xi}^2+p_{yi}^2+p_{zi}^2)/2m$. 
In the former statistical approach the phase space density of states (DoS) $\Omega$ is determined and then by means of the Boltzmann-Gibbs (BG) statistical entropic formula $S_\Omega=\kB \ln(\Omega)$ the thermodynamic expressions mentioned above are obtained.

However, in the recent study  \cite{OikGBB2013a}  it has been show that the limit $N\ra\infty$ is not an inherent thermodynamic property. In other words, the thermodynamic results are valid even for a finite number of molecules unveiling herewith that the physical Clausius and the statistical $S_\Omega$ entropies are distinct. The same conclusion is presented in Ref. \cite{Dieks} where the authors follow a different argumentation route based on  the particles (in)distinguishability. 
Not to forget that the phase space results  may reproduce the thermodynamic ones only for the specific choice of the constant $\alpha=2/3$. Although the former value can be justified within the kinetic theory it presents a restriction stemming from the phase space statistical model. 
The value of $\alpha$ is related to the time interval between two successive collisions with the (relevant) wall of the container. 
Generally, depending on the collisions of the molecules with  each other, the time interval between two successive collisions with the wall may be longer or shorter than the ideal one corresponding to $\alpha=2/3$. It is the physical system under scrutiny indicating the value of $\alpha$ and not a plugged by hand in an ad hoc manner.

In the current paper, continuing  the work started in Ref. \cite{OikGBB2013a}, we go a step beyond demonstrating that the difference  between the physical and the statistical entropies does not exclusively lie on the computation of the limit $N\rightarrow\infty$ and the related theoretical issues. Its origin is more general and attributed to the inappropriate of the BG statistical  representation of Clausius entropy $S$, namely $S=S_\text{DoS}\equiv\ln(\text{DoS})$.
For our purpose, in Section \ref{IGoADen} we explore thermodynamically ideal gases confined  in a finite container  for the general case of $\alpha>0$ and derive a new thermodynamic entropy expression, even in the dilute gas limit (infinite heat bath).
Herewith, preserving the BG-structure $S_\text{DoS}$, we determine the thermodynamic expression $\Phi$ corresponding to the DoS and compute in Section \ref{ThermoVsStat}  the associated partition function $Q^\Phi$   by virtue of the Laplace transformation of $\Phi$. Comparing the former  with the respective expressions obtained from the phase space statistics, $\Omega$ and $Q^\Omega$, we are led to a contradiction. Although both approaches describe indeed ideal gases their results are distinct.
A discussion on this discrepancy between phenomenological thermodynamics and thermostatistics follows.
Finally, in Section \ref{Conclusion} concluding remarks are presented.

%\newpage
%====================================================
\section{Ideal Gases in a finite heat bath}\label{IGoADen}
%====================================================
%
In this section we shall derive the basic thermodynamic expressions of ideal gases embedded  in a  finite heat bath. The total system is closed.
Our starting point  is the first thermodynamic law for nIaG in its differential form for reversible processes,  namely
%
%---------------------------------------------------------
\begin{equation}\label{FLT}
\text{d}S(E,V)=\frac{1}{T(E,V)}\text{d}E+\frac{\alpha\, E}{VT(E,V)}\text{d}V\,,
\end{equation}
%---------------------------------------------------------
%
where $E$ and $V$ are the independent variables and $S(E,V)$ is the Clausius entropy. 
From its integrability condition, i.e., $\frac{\partial^2S}{\partial V \partial E}=\frac{\partial^2S}{\partial E \partial V}$, the following partial differential equation  is obtained
%
%---------------------------------------------------------
\begin{equation}\label{TempCond}
E\frac{\partial T(E,V)}{\partial E}-\frac{V}{\alpha}\frac{\partial T(E,V)}{\partial V}=T(E,V)\,.
\end{equation}
%---------------------------------------------------------
%
The general solution of Eq. \eqref{TempCond} has the form
%
%---------------------------------------------------------
\begin{equation}\label{GenSolTemp}
T(E,V)=E\,f\rbr{E^{1/\alpha}V}\,.
\end{equation}
%---------------------------------------------------------
%
$f$ is an arbitrary smooth function.
Eq. \eqref{GenSolTemp} unveils that nIaG's are mathematically compatible with a more complex temperature behaviour with respect to the averaged internal kinetic energy rather than the linearity in Eq. \eqref{HeatCap} ($f$ being a constant function) and  they may depend on the volume of the system as well.
Assuming bijectivity between $T$ and $E$ Eq. \eqref{FLT} can be rewritten as
%
%---------------------------------------------------------
\begin{equation}\label{FLT_U}
\text{d}S(T,V)=\frac{1}{T}\frac{\partial E(T,V)}{\partial T}\text{d}T
+\frac{1}{T}\sqbr{\frac{\partial E(T,V)}{\partial V}+\frac{\alpha\,E(T,V)}{V}}\text{d}V\,.
\end{equation}
%---------------------------------------------------------
%
The integrability condition of Eq. \eqref{FLT_U} gives
%
%---------------------------------------------------------
\begin{equation}\label{EnergyCond}
T\frac{\partial E(T,V)}{\partial T}
-\frac{V}{\alpha}\frac{\partial E(T,V)}{\partial V}
=E(T,V)\,,
\end{equation}
%---------------------------------------------------------
%
yielding
%
%---------------------------------------------------------
\begin{equation}\label{GenSolEnergy}
E(T,V)=T\,g\rbr{T^{1/\alpha}V}\,.
\end{equation}
%---------------------------------------------------------
%
Similar to $f$, $g$ is an arbitrary smooth function.
Computing then, the partial derivative of $E$ in Eq. \eqref{GenSolEnergy} with respect to the temperature we obtain
%
%---------------------------------------------------------
\begin{equation}\label{heatCap}
g(x)+\frac{x}{\alpha}\,\frac{dg(x)}{dx}=C_V(T)\,,
\end{equation}
%---------------------------------------------------------
%
where $x\equiv T^{1/\alpha}V$.
We remark that Eq. \eqref{heatCap} is valid for gases comprised of noninteracting molecules. The choice then, of the form of $C_V(T)$ corresponding to the physical situation under scrutiny, forms the respective structure of $g(x)$.
%Comparing Eqs. \eqref{bijectivity} and \eqref{heatCap}, we see that the nIaG constant thermal capacity may be positive as well as negative.
%
%For the sake of simplicity we will consider further only $C_V>0$.
%

Requesting a constant $C_V$, the former differential equation is solved as
%
%---------------------------------------------------------
\begin{equation}\label{det_g}
g(x)=C_V-\vartheta\,x^{-\alpha}\,,
\end{equation}
%---------------------------------------------------------
%
where $\vartheta$ is an integration constant.
%
%From Eqs. \eqref{rel_f_g} and \eqref{det_g} we determine the function $f$ as well, i.e., $f(x)=(1+\vartheta\,x^{-\alpha})/C_V$. 
%
Herewith, the relation between $E$ and $T$ is determined as
%
%---------------------------------------------------------
\begin{equation}\label{Energy_Temp}
E(T,V)=C_V\,T-\vartheta\,V^{-\alpha}
\qquad\Leftrightarrow\qquad
T(E,V)=\frac{1}{C_V}\sqbr{E+\vartheta\,V^{-\alpha}}\,.
\end{equation}
%---------------------------------------------------------
%
As can be seen, the averaged internal kinetic energy   for nIaG with constant $C_V$  presents a linear dependence on the temperature as much as a power law dependence on the volume of the container.
%
%From Eqs. \eqref{HeatCap} and \eqref{Energy_Temp} we read the analytical expression of $\Theta$
%
%---------------------------------------------------------
%\begin{equation}\label{A}
%\Theta(V)=-\vartheta\,V^{-\alpha}\,.
%\end{equation}
%---------------------------------------------------------
%
%Evidently, $\Theta$ satisfies the limit $\lim_{V\rightarrow\infty}\Theta(V)=0$ recovering the expressions of gases in the law pressure regime. Accordingly, the information of the finiteness of the heat bath is included in the expression of  the $\Theta$ function.
%
%Its dimension is of the energy so that the  constant $\vartheta$ has the dimension of $J/m^{3\alpha}$.
%
%
Plugging  Eq. \eqref{Energy_Temp} into Eq. \eqref{eq:intro-01} we obtain the respective equation of state, namely
%
%---------------------------------------------------------
\begin{equation}\label{newEqOfSt}
PV=\alpha C_VT-\alpha \vartheta V^{-\alpha}\,,
\end{equation}
%---------------------------------------------------------
%
and a  straightforward integration of  Eq. \eqref{FLT} yields the Clausius entropy for our class of gases
%
%---------------------------------------------------------
\begin{equation}\label{Entropy10}
S(E,V)=C_V \ln(\vartheta+E V^\alpha)+d \qquad\Leftrightarrow\qquad
S(T,V)=C_V \ln(C_V T V^\alpha)+d\,,
\end{equation}
%---------------------------------------------------------
%
where $d$ is the integration constant and thus it may present a dependence on $C_V$.
Due to the logarithmic domain we have $C_V T>0$.
As can be seen, when $V\rightarrow\infty$ Eqs. \eqref{Energy_Temp} - \eqref{Entropy10} recover the analogous expressions of the dilute ideal gases with $C_V=nR/\alpha$. 
For a given number $N$ of molecules the former condition is equivalent to the DiG limit $N/V\rightarrow 0$. 
It becomes then obvious, that the aforementioned equations describe ideal gases confined in a finite thermal bath.

\begin{figure}[t]
\begin{center}  
  % Requires \usepackage{graphicx}
    \includegraphics[width=8.1cm,height=6.6cm]{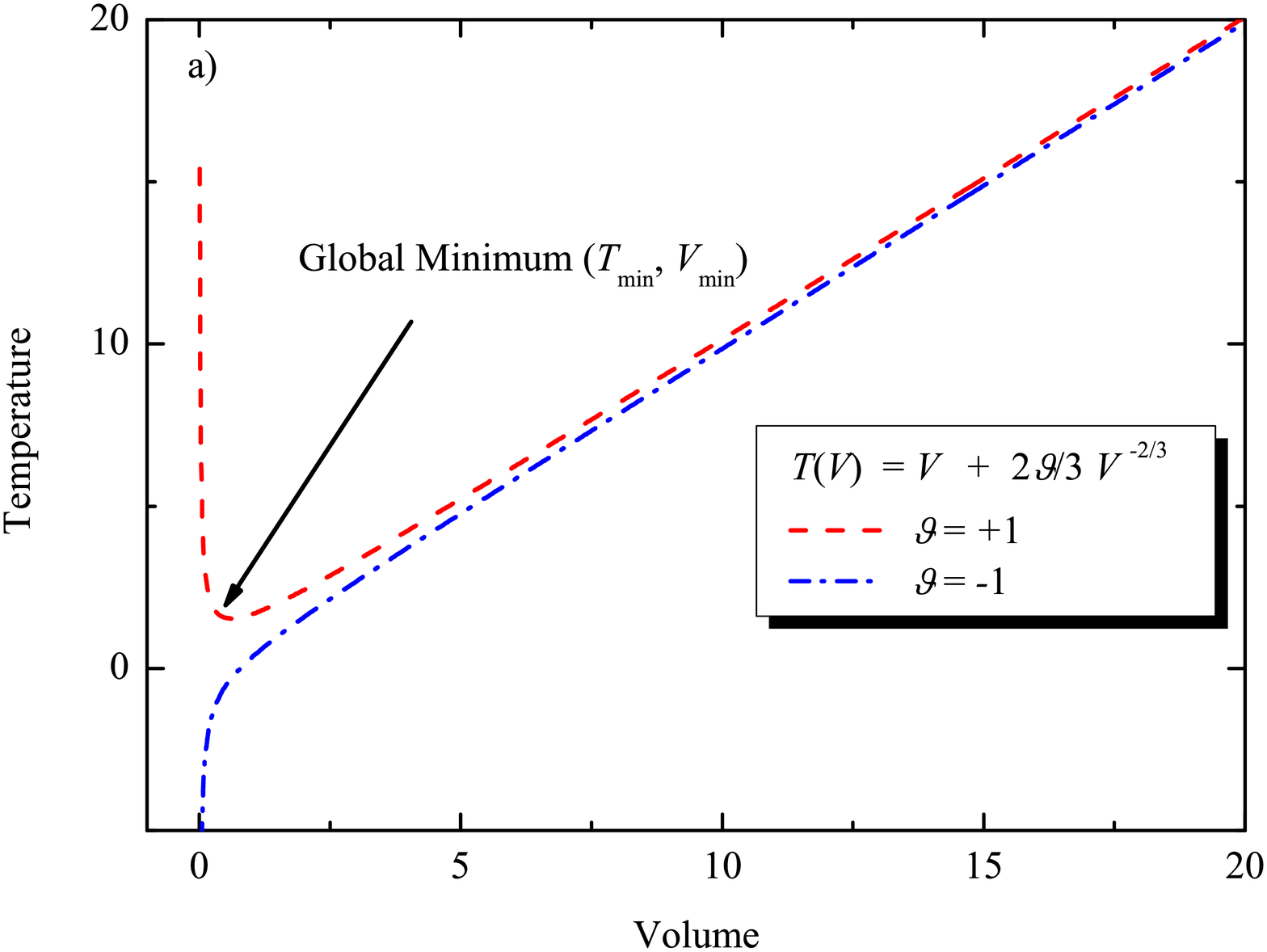}
      \includegraphics[width=8.1cm,height=6.6cm]{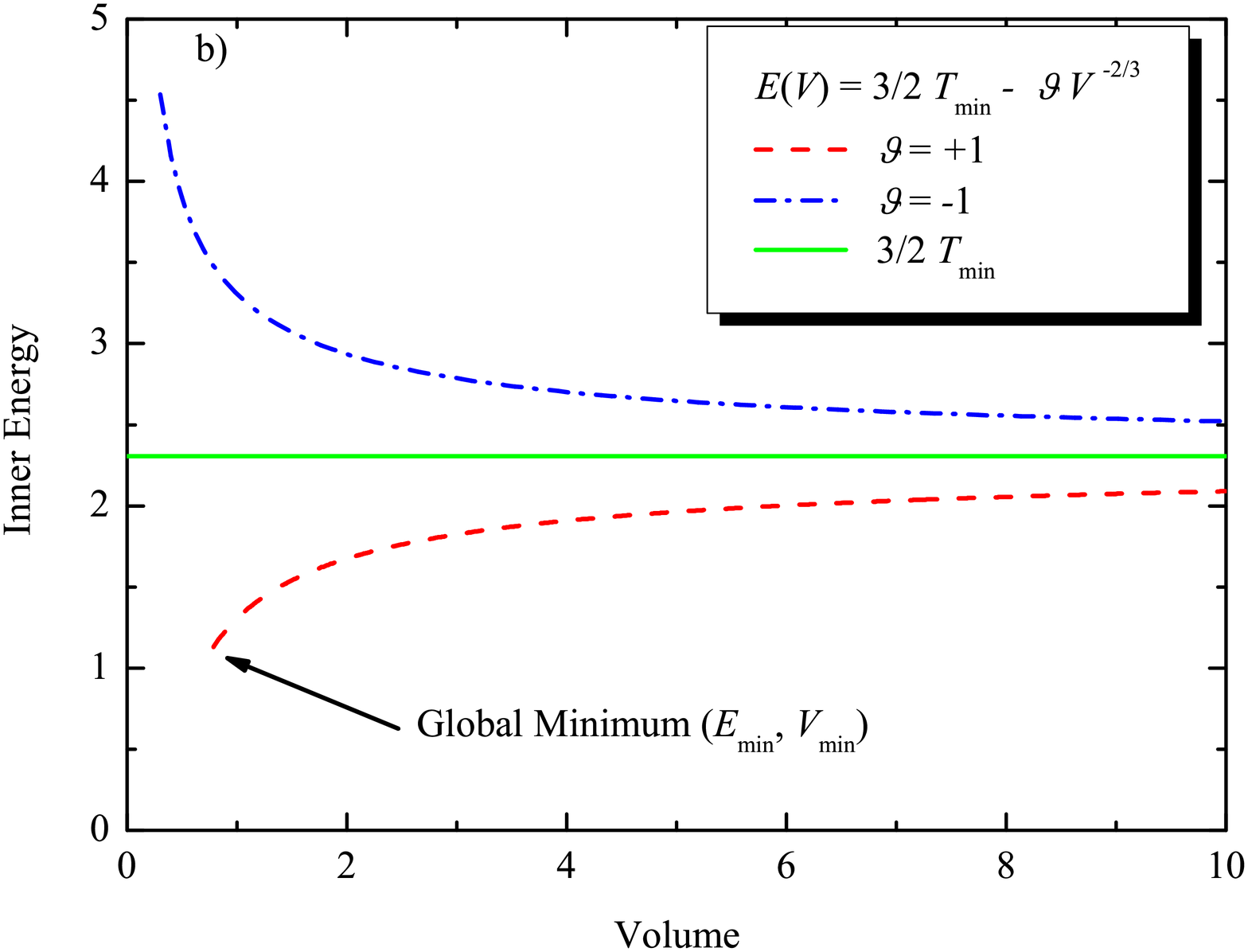}
    \caption{ Plot of a) temperature and b) energy depending on the container volume $V$  for $C_V=1/\alpha$,  $\alpha=2/3$, $P=1$ and $\vartheta=\pm1$.}%
    \label{Temp_Vol_fig}
\end{center}
\end{figure}
In Fig. \ref{Temp_Vol_fig}a) we plot the temperature of Eq. \eqref{newEqOfSt}  as a function of the container volume for $C_V=1/\alpha$, $\alpha=2/3$, $P=1$ and $\vartheta=\pm1$, i.e., $T(V)=V\pm\frac{2}{3}V^{-2/3}$.
As can be seen, for large values of the volume the linear term dominates, corresponding to the dilute ideal gas case, while for small volume values the leading term is the power law  causing deviations from linearity. 
In the former regime, we observe two different temperature behaviours depending on the sign of $\vartheta$. 
For $\vartheta<0$, see in Fig. \ref{Temp_Vol_fig}b), the energy $E$  increases for decreasing volume.
The former energy behaviour however is not consistent with the statistical assumptions of Eq. \eqref{eq:intro-01} in its entire domain, since by reducing $V$ the collisions will become more frequent and stronger so that the molecule interactions may not be further negligible.
When $\vartheta$ is positive on the other hand, see Figs. \ref{Temp_Vol_fig}a) and \ref{Temp_Vol_fig}b), the  temperature in Eq. \eqref{newEqOfSt} presents a finite global minimum corresponding to a positive minimum energy value, i.e.,
%
%---------------------------------------------------------
\begin{equation}\label{MinVal}
V_{\min}=\rbr{\frac{\vartheta\,\alpha^2}{P}}^{\frac{1}{\alpha+1}}\,,\qquad
T_{\min}=\frac{\alpha+1}{\alpha C_V} 
\rbr{\frac{\vartheta\,P^\alpha}{\alpha^{\alpha-1}}}^{\frac{1}{\alpha+1}}\,,\qquad
E_{\min}=\rbr{\frac{\vartheta\,P^\alpha}{\alpha^{\alpha-1}}}^{\frac{1}{\alpha+1}}\,.
\end{equation}
%---------------------------------------------------------
%
%
Below this point the behaviour is not of physical relevance, since a further  increase of temperature corresponds to a decrease of the volume. 
%
%We remark that by changing the notation $T_{\min}\ra T_{\max}$ the expressions in Eq. \eqref{MinVal} are valid for $C_V<0$ as well. Accordingly, the internal energy $E$ of the gas is is always positive.
%
Combining all the relations in Eq. \eqref{MinVal} we can express $\vartheta$ in terms of $T_{\min},V_{\min}$ and $E_{\min}$ as
%
%---------------------------------------------------------
\begin{equation}\label{subsecB_01}
\vartheta=\frac{C_V}{(\alpha+1)}T_{\min}V_{\min}^{\alpha}=\frac{1}{\alpha}E_{\min}V_{\min}^{\alpha}\,.
\end{equation}
%---------------------------------------------------------
%
In this case the reduction of the internal energy when volume decreases  preserves now the negligibility of the molecule interactions, in contrast to the case of the negative $\vartheta$, as discussed above.
Therefore, in what follows we consider $\vartheta>0$.

Taking Eq. \eqref{subsecB_01} into account, Clausius entropy can be rewritten as
%
%-----------------------------------------------------------------------------------------
\begin{equation}\label{newEntropy}
S(E,V)=C_V \ln\big[1+W(E,V)\big]+C_V\ln(q)\,,
\end{equation}
%-----------------------------------------------------------------------------------------
%
where $W(E,V):=\alpha\rbr{\frac{E}{E_{\min}}}\rbr{\frac{V}{V_{\min}}}^\alpha$ and $q:=\vartheta\exp(d/C_V)$.
We stress here that $E_{\min}$ and $V_{\min}$ are additive quantities and not just some arbitrarily chosen initial values, as one may think at first sight. Thus, a doubling of the system, for example, would change their values to $2E_{\min}$ and $2V_{\min}$ , respectively. Accordingly, the function $W$ is intensive, $W (\lambda\,E, \lambda\,V ) = W (E, V )$, and choosing $d$ in such a way that $q$ is a constant, then Clausius entropy becomes extensive. 
%
%Considering $S$ in the limit $V \rightarrow\infty$ (the unity in the logarithmic term becomes negligible) we obtain the thermodynamic entropy for DiIG. 
%
The result in Eq. \eqref{newEntropy} to the best of our knowledge presents a new expression in literature of the Clausius entropy even for dilute ideal gases, in which although the logarithmic argument does not explicitly depend on $n$ (or $C_V$ or $N$ ), $S$ may satisfy the extensivity property. It is worth noting that the former expression is obtained only under the complete description of ideal gases, i.e., considering the finiteness of the heat bath. 
Indeed, studying them  only in the DiG limit we miss the information of $\{E_{\min} , V_{\min} \}$ so that any effort to obtain extensivity in the logarithm has been performed by means of $n$-dependent terms.

Herewith, we have derived all relevant for our purpose thermodynamic equations for the class of gases under scrutiny. In the next section we will attempt to reproduce the former within the phase space statistical approach and discuss the results.
\section{The Boltzmann-Gibbs entropic formula}\label{ThermoVsStat}
%==================================================
%
Let us first give a brief review of the necessary for our analysis  phase space statistical equations for nIaG.
The energy probability distribution $\rho(E)$  of noninteracting molecules is of exponential form, i.e., $\rho(E)=\exp(-\beta E)/Q$. $Q$ is the partition (normalization) function computed in the continuous limit as $Q=Q_1^\text{DoS}=Q_2$ with $Q^{\mathrm{DoS}}_1\equiv\int_0^\infty e^{-\beta E}\mathrm{DoS}(E)\mathrm{d}E$ and $Q_2\equiv(N!h^{3N})^{-1}\int_\mathbb{R} e^{-\beta H(\mathbf{p})} \mathrm{d}^{N} \mathbf{q} \mathrm{d}^{N} \mathbf{p}$. $H(\mathbf{p})$ is the hamiltonian comprised of kinetic energy terms and $h$ is the Planck constant.
At this stage $\beta$ is solely an energy factor.
From the latter integral we determine the analytical expression of $Q$
%
%-----------------------------------------------------------------------------------
\begin{equation}\label{ParFun3}
Q_2(\beta,V)=\frac{1}{N! h^{3N}}\int_\mathbb{R} e^{-\beta H(\mathbf{p})} \mathrm{d}^N\mathbf{q}\mathrm{d}^N\mathbf{p}= 
\frac{1}{N!}\rbr{\frac{V}{h^3}}^N \rbr{\frac{2\pi m}{\beta}}^{3N/2}\,.
\end{equation}
%-----------------------------------------------------------------------------------
%
%
Then, the phase space density of states $\Omega$ can be estimated by virtue of the inverse Laplace transformation of $Q_1^{\text{DoS}}$, i.e.,
%
%-----------------------------------------------------------------------------------
\begin{equation}\label{Eq00A1}
\Omega(E,V)=\frac{1}{2\pi \mathrm{i}}\lim_{L\ra\infty} \int_{\beta'-\mathrm{i} L}^{\beta'+\mathrm{i} L} e^{\beta E}Q^\Omega_1(\beta,V)\mathrm{d}\beta
\end{equation}
%-----------------------------------------------------------------------------------
%
under the identification $Q^{\mathrm{DoS}}_1(\beta,V)=Q_2(\beta,V)$, yielding  \cite{Pathria}
%
%-----------------------------------------------------------------------------------
\begin{equation}\label{PSDensOfSt2}
\Omega(E,V)=\frac{V^N}{N!}\rbr{\frac{2\pi m}{h^2}}^{3N/2} \frac{E^{(3N/2)-1}}{\{(3N/2)-1\}!}\,.
\end{equation}
%-----------------------------------------------------------------------------------
%
%
The relation between $\beta$ and $E$ is determined from the equipartition theorem of the canonical ensemble, namely
%
%-----------------------------------------------------------------------------------
\begin{equation}\label{equipart}
\av{p_i \frac{\partial H(\mathbf{p})}{\partial p_j}}=\frac{\int \rbr{p_i \frac{\partial H(\mathbf{p})}{\partial p_j}} e^{-\beta H(\mathbf{p})} \mathrm{d}^{N}\mathbf{q}\mathrm{d}^{N}\mathbf{p}}{\int e^{-\beta H(\mathbf{p})} \mathrm{d}^{N}\mathbf{q}\mathrm{d}^{N}\mathbf{p}}
\qquad\Rightarrow\qquad
\beta E=\frac{3}{2}N\,.
\end{equation}
%-----------------------------------------------------------------------------------
%
Eqs. \eqref{ParFun3}-\eqref{equipart} present general statistical results of the phase space approach for nIaG without invoking any thermodynamic ingredients. 
The connection now between thermodynamics and statistics for the former class of gases is succeeded through the analytical structure of $\beta$. Indeed,  comparing Eq. \eqref{equipart} with Eq. \eqref{GenSolEnergy} we obtain
%
%-----------------------------------------------------------------------------------
\begin{equation}\label{betaFac0}
\beta=\frac{3N}{2Tg(T^{1/\alpha}V)}\,.
\end{equation}
%-----------------------------------------------------------------------------------
%
Here, we see the important role of the factor $\beta$ in the above equations, namely  it contains the entire thermodynamic information about nIaG in the reservoir.
For the specific case of ideal gases in a finite reservoir $\beta$ reduces to
%
%-----------------------------------------------------------------------------------
\begin{equation}\label{betaFac}
\beta=\frac{3N}{2\big(C_V T-\vartheta V^{-\alpha}\big)}\,,
\end{equation}
%-----------------------------------------------------------------------------------
%
and it recovers the textbook expression $\beta^{-1}=\kB T$ in the limit $V\ra\infty$ for $C_V=nR/\alpha$ and $\alpha=2/3$.

Returning to thermodynamics and expressing the Clausius entropy in Eq. \eqref{newEntropy} in  the BG form $S=\kB \ln(\mathrm{DoS})$,  we have
%
%-----------------------------------------------------------------------------------
\begin{equation}\label{PSDensOfSt1}
\Phi(E,V)=\sqbr{q\rbr{1+\frac{EV^\alpha}{\vartheta}}}^{N/\alpha}\,,
\end{equation}
%-----------------------------------------------------------------------------------
%
where $\Phi$ corresponds to the phase space density of states obtained within the phenomenological thermodynamics.
We remind that $q$ is a constant due to the extensivity of $S$ and the constant $\vartheta$ may be given as in Eq. \eqref{subsecB_01}.
Comparing now Eqs. \eqref{PSDensOfSt2} and \eqref{PSDensOfSt1}, we see that although both results are exact and describe both correctly the ideal gases (they are valid as much for finite $N$ as for finite $V$ and any $\alpha$), we have $\Omega\neq \Phi$. This is apparently a contradiction.
These  two expressions may merely coincide when considering an infinite bath  with infinite number of molecules subject the proper choice of the constants. 
Computing the associated to $\Phi$ partition function  $Q^\Phi_1$ we obtain
%
%-----------------------------------------------------------------------------------
\begin{equation}\label{ParFun}
Q^\Phi_1(\beta,V)=\int_0^{\infty} e^{-\beta E}\Phi(E)\mathrm{d}E
=\rbr{\frac{q}{\vartheta}}^{N/\alpha}\frac{V^N}{\beta^{N/\alpha+1}}\,\, e^{\beta \vartheta V^{-\alpha}}\,\Gamma\rbr{\frac{N}{\alpha}+1,\beta\vartheta
 V^{-\alpha}}\,.
\end{equation}
%-----------------------------------------------------------------------------------
%
Obviously, the partition functions $Q_1^\Omega$ and $Q_1^\Phi$ in Eqs. \eqref{ParFun3} and \eqref{ParFun} are distinct.
%, respectively, are distinct and one can easily verify that they do not coincide even for $N,V\ra \infty$ and $\alpha=2/3$. Here, it becomes evident that the coincidence in the expressions of $\Phi$ and $\Omega$ in the aforementioned limits is of purely mathematical nature.
%
Then, the difference in the expressions of the statistical $\Omega$ and the thermodynamical $\Phi$ density of states, preserving the BG-structure $S=\kB \ln(\text{DoS})$, may be attributed to the  non-equality 
%
%-----------------------------------------------------------------------------------
\begin{equation}\label{NonEquiv}
\int_0^\infty e^{-\beta E}\mathrm{DoS}(E)\mathrm{d}E \neq \frac{1}{N!h^{3N}}\int_\mathbb{R} e^{-\beta H(\mathbf{p})} \mathrm{d}^{N}\mathbf{q} \mathrm{d}^N\mathbf{p}
\qquad\Leftrightarrow\qquad Q_1^\mathrm{DoS}\neq Q_2\,.
\end{equation}
%-----------------------------------------------------------------------------------
%
Eq. \eqref{NonEquiv} gives the following information. If ones assumes that $Q_1^\text{DoS}$ is a meaningful passage from the discrete  $Q_1^\text{DoS}=\sum_{\ell(levels)}\text{DoS}(E_\ell)e^{-\beta E_\ell}$ to the continuous partition function, then the correspondence between phase space coordinates integration and energy integration must be revisited and properly modified, so that from $Q_2$ one should be able to obtain Eq. \eqref{ParFun}.
Assuming the correctness of $Q_2$ on the other hand,  then a proper continuous form of $Q_1^\text{DoS}=\sum_{\ell(levels)}\text{DoS}(E_\ell)e^{-\beta E_\ell}$ needs to be explored being in agreement with the thermodynamic correspondence of the density of states, i.e, DoS$(E,V$)$ =\Phi(E,V)$ in Eq. \eqref{PSDensOfSt1}.  
However, in both cases any change in the above integrals of the partition function $Q$ would cause a drastic and nontrivial change in the correspondence between sums and integrals throughout the statistical and physical theories.
Therefore, in author's opinion, the most natural explanation of the result $\Phi\neq\Omega$ is to preserve the equivalence between $Q_1^\text{DoS}$ and $Q_2$ and accept that the BG structure proposal for a statistical representation of the Clausius entropy is not suitable. 
In this case, $\Omega$ in Eq. \eqref{PSDensOfSt2} is indeed the density of states of ideal gases (and even of nIaG), yet the phase space measure  $\ln(\Omega)$ does not represent the physical entropy in contrast to $\ln(\Phi)$. 

%Moreover, this alternative is in agreement with the Boltzmann derivation of the exponential energy distribution, i.e., logarithmic expansion of the probability \textcolor{red}{$P_r\sim\Omega(E_{tot}-E_r)$, namely}
%
%-----------------------------------------------------------------------------------
%\begin{equation}\label{betaDef}
%\beta:=\frac{\partial \ln(\Omega)}{\partial E}\bigg|_{E=E^*}
%\overset{\eqref{Energy_Temp}}{=}\frac{3N}{2}
%\frac{1}{C_V T-\vartheta V^{-\alpha}}\,.
%\end{equation}
%-----------------------------------------------------------------------------------
%
%
%Indeed, the above definition of $\beta$ yields  the same result as in Eq. \eqref{betaFac}, where $\Omega$ gives the density of states over the phase space volume $\Omega$ \cite{Scheck} instead of the phase space density $\Omega$, $\Omega=\partial \Omega/\partial E$. 
%
%The appropriateness of the definition in \eqref{betaDef}  over $\Omega$ can be verified directly from the self-consistency of $\Omega$, independent from the thermodynamic result in Eq. \eqref{Energy_Temp}.
%
%Plugging $E$ from Eq. \eqref{equipart} into Eq. \eqref{PSDensOfSt2} and then replacing $\beta$ as $\partial \ln(\Omega\,\text{or}\,\Omega)/\partial E$, we immediately see that $\Omega$ remains unaltered only under the definition in Eq. \eqref{betaDef}.

%\newpage

%==================================================
\section{Conclusions}\label{Conclusion}
%==================================================
%
Summarizing, we have derived the basic thermodynamic relations for  ideal gases embedded in a finite heat bath (container).
We showed that the respective internal kinetic energy presents a linear dependence on the temperature and a power law dependence on the volume of the container. The latter dependence vanishes in the dilute gas limit (infinite heat bath), recovering the structure of the known textbook expressions of the dilute ideal gases. 
%
%Furthermore, our approach unveiled that thermodynamics of ideal gases is compatible with both positive and negative temperatures corresponding simultaneously to positive and negative thermal capacities, respectively, while the energy of the gas is always positive.
%
We have then determined within phenomenological  thermodynamics, preserving the structure of the Boltzmann-Gibbs entropic formula, the physical correspondence $\Phi$ to the phase space density of states for our class of gases and computed the respective partition function $Q_\Phi$ of the canonical ensemble by means of the Laplace transformation of $\Phi$.
We observed that the former derived expressions are distinct from the analogous ones obtained within phase space statistics. 
The aforementioned discrepancy is attributed either to the non-equivalence between the energy integral and  the corresponding integration over the phase space coordinates in  the partition function or to the invalidity of the Boltzmann-Gibbs representation of the Clausius entropy, i.e.,  $S=\ln(Density \,\,of\,\,States)$. We argued for the appropriateness of the latter choice. 

%Finally, we remarked that our results are exact and self-consistent under the identification of the phase space density of states via the phase space volume rather than the phase space surface.

%============================
\section*{Acknowledgments}
%============================
\noindent
This work has been supported by TUBITAK (Turkish Agency) under the Research Project number 112T083, by the THALES Project MACOMSYS, funded by the ESPA Program of the Ministry of Education of Hellas, and by the European Union's Seventh Framework Programme (FP7-REGPOT-2012-2013-1) under grant agreement n$^{\text{o}}$ 316165.

%============================

\end{document}